# Basis-neutral Hilbert-space analyzers

Lane Martin[1], Davood Mardani[2], H. Esat Kondakci[1], Walker D. Larson[1], Soroush Shabahang[1], Ali K. Jahromi[1], Tanya Malhotra[3,4], A. Nick Vamivakas[4,5], George K. Atia[2], and Ayman F. Abouraddy[1*]

[1]*CREOL, The College of Optics & Photonics, University of Central Florida, Orlando, FL 32816, USA*

[2]*Dept. Electrical Engineering and Computer Engineering, University of Central Florida, Orlando, FL 32816, USA*

[3]*Department of Physics and Astronomy, University of Rochester, Rochester, New York 14627, USA*

[4]*Center for Coherence and Quantum Optics, University of Rochester, Rochester, New York 14627, USA*

[5]*Institute of Optics, University of Rochester, Rochester, NY 14627, USA*

*Corresponding author; email: raddy@creol.ucf.edu

**Interferometry is one of the central organizing principles of optics. Key to interferometry is the concept of optical delay, which facilitates spectral analysis in terms of time-harmonics. In contrast, when analyzing a beam in a Hilbert space spanned by spatial modes – a critical task for spatial-mode multiplexing and quantum communication – basis-specific principles are invoked that are altogether distinct from that of 'delay'. Here, we extend the traditional concept of *temporal* delay to the *spatial* domain, thereby enabling the analysis of a beam in an arbitrary spatial-mode basis – exemplified using Hermite-Gaussian and radial Laguerre-Gaussian modes. Such generalized delays correspond to optical implementations of fractional transforms; for example, the fractional Hankel transform is the generalized delay associated with the space of Laguerre-Gaussian modes, and an interferometer incorporating such a 'delay' obtains modal weights in the associated Hilbert space. By implementing an inherently stable, reconfigurable spatial-light-modulator-based polarization-interferometer, we have constructed a 'Hilbert-space analyzer' capable of projecting optical beams onto any modal basis.**

Interferometry is the cornerstone of fundamental investigations and precise measurements in optics[1]. The nature of light – both classical[2,3] and quantum[4-6] – was unraveled largely through interferometric experiments, and the exquisite precision inherent in optical interferometry has been instrumental in metrology[7], bio-imaging[8], devising ultra-sensitive systems for the detection of gravitational waves[9], and enabling novel lithographic schemes[10]. These examples share a common feature: interference results from combining beams with relative phases engendered by optical delays. A principal utility for optical interferometry is spectral analysis – determining the contributions of the *continuum* of time-frequency harmonics to the optical signal. Recent applications have emphasized the utility of *discrete* spatial-mode bases for optical beams, such as orbital angular momentum (OAM) states[11-13] exploited in free-space[14,15] and multimode fibers[16,17] to increase their information-carrying capacity (so-called spatial-mode multiplexing) and in quantum communication protocols[18] (such as quantum key distribution[19]). An optical beam in this conception is an element in a Hilbert space spanned by such a basis. In general, strategies for spatial-mode analysis rely on approaches altogether different from the concept of optical delays that has served interferometry so well. In other words, we currently lack a 'Hilbert-space analyzer': a hypothetical device capable of analyzing an optical beam in the vector space defined by any prescribed modal basis. Examples of strategies for modal analysis range from phase-retrieval combined with direct mode projections[20], correlating the modes with spectral or temporal degrees of freedom[21], combining principal-component analysis after adapting the detection system with a training data set[22], to performing a coordinate transformation that converts the beam into a more convenient basis[23]. In particular, despite multiple techniques for OAM beam analysis[24-27], comparable progress has been lacking for other important modal bases, such as radial Laguerre-Gaussian[28-31] (LG) modes.

In archetypal two-path interferometers, two copies of a beam are combined after a relative optical delay is inserted. The delay is swept and an interferogram is traced, which yields the modal weights of time-frequency harmonics through spectral analysis[32]. In this paper, we present a unifying principle for modal analysis by addressing the following question: can the traditional optical delay – one of the most fundamental concepts in optics – be extended beyond its implementation in the time domain to apply to Hilbert spaces associated with discrete spatial-mode bases? We show here that such a generalization is indeed possible. We introduce the concept of a *generalized delay* (GD): an optical transformation characterized by a continuous, real order-parameter that can be tuned to produce – once placed in one arm of an interferometer – an interferogram that reveals the modal weights in a prescribed functional basis via harmonic analysis. We find that GDs correspond to optical implementations of fractional transforms in the case of discrete modal bases[33,34]. For example, it can be shown[34] that the GD associated with Hermite Gaussian (HG) modes is the fractional Fourier transform[35-36], whereas that associated with radial LG modes[33] is the fractional Hankel transform[37,38]. Sweeping the order of a fractional transform corresponds to varying a temporal delay in traditional interferometry – each in its own Hilbert space.

In the implementation presented here, we exploit electrically addressable spatial light modulators (SLMs) to realize tunable-strength cylindrical and spherical lenses that are building blocks of fractional transforms[39]. We make use of the polarization discrimination of SLMs[40] to construct a polarization interferometer – in lieu of a two-path interferometer – to accomplish generalized interferometry in an inherently stable configuration. Switching between Hilbert spaces – that is, examining a beam in different bases – is readily achieved in the same setup with no moving parts, simply by changing the phases imparted by the SLMs. We thus establish a versatile, basis-neutral Hilbert-space analyzer based on a generalized conception of optical interferometry.

**Concept of a generalized optical delay.** An optical delay $\tau$ is typically implemented by inserting an additional propagation length in a beam's path. In the time domain, a delay shifts the temporal origin $E(t) \rightarrow E(t-\tau)$, whereas in the spectral domain it adds to each harmonic frequency component $\omega$ a phase $e^{i\omega\tau}$ that is linear in both the delay and the frequency (Fig. 1a). In other words, spectral harmonics $\{e^{-i\omega t}\}$ are eigenstates of the delay operation with eigenvalues $e^{i\omega\tau}$. Guided by this observation, we introduce a generalized delay (GD) that operates in the Hilbert space spanned by a modal basis $\{\psi_n(x)\}$, such that the GD's effect on a beam described in this space is completely analogous to that of a delay $\tau$ for a pulse. A GD operates between an input plane $x'$ and output plane $x$, and implements a unitary transformation $\Lambda(x, x'; \alpha)$ characterized by a real, continuous order-parameter $\alpha$,

$$\Lambda(x, x'; \alpha) = \sum_n e^{in\alpha} \psi_n(x)\psi_n^*(x'), \qquad (1)$$

where the functional basis $\{\psi_n(x)\}$ is orthonormal and complete, and its members are the eigenstates of $\Lambda$: they emerge from the GD unchanged except for a mode-dependent phase $e^{in\alpha}$ (Fig. 1b), $\int dx' \Lambda(x, x'; \alpha)\psi_n(x') = e^{in\alpha}\psi_n(x)$; see Methods.

Consider a monochromatic beam $E(x) = \sum_n c_n \psi_n(x)$, where $\{c_n\}$ are modal coefficients and $E(x)$ is normalized $\int dx |E(x)|^2 = 1$, such that $\sum_n |c_n|^2 = 1$. Upon passage through the GD, the field is transformed according to

$$E(x; \alpha) = \int dx \Lambda(x, x'; \alpha) E(x') = \sum_n c_n e^{in\alpha} \psi_n(x). \qquad (2)$$

Each mode thus acquires a phase $e^{in\alpha}$ that depends linearly on its index $n$ (Fig. 1b) – in analogy to the impact of a traditional delay with respect to spectral harmonics. For a discrete modal basis indexed by $n$ (Eq. 1), $\Lambda(x, x'; \alpha)$ is periodic in $\alpha$ with period $2\pi$. Furthermore, $\Lambda$ can be generalized to two transverse coordinates and is applicable to a continuous basis[33,34].

As an example, consider the set of one-dimensional (1D) HG modes, $H_n(x) = A_n e^{-x^2} h_n(x)$, where $h_n(x)$ is the $n^{\text{th}}$-order Hermite polynomial and $A_n$ is a normalization constant. This modal set is well-established as a useful basis for laser beams and arises naturally in many contexts[41]. The corresponding GD is the 1D fractional Fourier transform (fFT)[34] of angular order $\alpha$ (scaled heretofore by convention from 0 to 4). Indeed, HG modes are eigenstates of the fFT[36] with eigenvalues $e^{i\pi n\alpha/2}$. A beam traversing this GD is not shifted in *physical* space, as an optical delay shifts a pulse in time. Nevertheless, because each underlying HG mode acquires the requisite phase after the GD, the fFT 'delays' the beam *in the Hilbert space of optical beams spanned by HG modes*, which thus facilitates analyzing the beam in the HG basis. Alternatively the set of *radial* LG modes associated with zero-OAM states given by $\phi_n(r) = B_n e^{-r^2/2} L_n(r^2)$ constitutes a modal basis for radial functions having azimuthal symmetry; here $L_n(\cdot)$ is the $n^{\text{th}}$-order Laguerre polynomial, $B_n$ is a normalization constant, and $r$ is a radial coordinate. The GD here corresponds to the fractional Hankel transform (fHT); i.e., an optical implementation of the fHT 'delays' the beam in the Hilbert space spanned by radial LG modes[33]. Techniques for beam analysis into radial LG modes are lacking, leading the radial coordinate to be recently dubbed 'the forgotten degree of freedom'[29].

**Generalized optical interferometry.** A GD can be exploited for the modal decomposition of an optical beam in its associated Hilbert space. The overall scheme for 'generalized optical interferometry' is a balanced two-path interferometer, in which the usual temporal delay is replaced by a GD (Fig. 2a). For an incident beam $E(x) = \sum_n c_n \psi_n(x)$ and a GD constructed using the modal basis $\{\psi_n(x)\}$, the output field is $E_t(x; \alpha) \propto \sum_n c_n (1 + e^{in\alpha}) \psi_n(x)$ and the power recorded by a 'bucket detector' is

$$P(\alpha) \propto \int dx |E_t(x; \alpha)|^2 \propto 1 + \sum_n |c_n|^2 \cos n\alpha, \qquad (3)$$

such that harmonic analysis of $P(\alpha)$ identifies the weights $|c_n|^2$; Fig. 2b. Each mode thus produces *individually* a sinusoidal interferogram $\propto 1 + \cos n\alpha$. Mode-orthogonality dictates that each mode interferes only with itself. Crucially, the form of the interferogram in Eq. 3 is *independent* of the particular modal basis. A superposition of two HG modes of order $n$ and $m$, for example, yields an interferogram that is identical to the same superposition of LG modes of order $n$ and $m$ – if the appropriate GD associated with each Hilbert space is implemented. This generalized interferometer is thus 'basis-neutral'. Furthermore, since the GD associated with a discrete modal basis is periodic in its order $\alpha$, the resulting interferogram is in turn periodic, such that its Fourier transform yields a discrete spectrum. The number of modes that may be distinguished in this manner is determined by the sampling rate of the interferogram (the number of settings of $\alpha$ measured) and is ultimately Nyquist-limited.

**Experimental implementation**. A fFT or fHT can be implemented via combinations of cylindrical or spherical lenses, respectively, and the transform orders are varied by changing either the lens strengths or their separation (or both)[36,42]. The former approach does not require moving parts and can be realized with electrically addressable phase-only SLMs that implement generalized lenses of variable power – which is the strategy we follow here. A minimum of three generalized lenses can implement a 1D fFT[39], where the first and last lenses have the same power and the distances separating the SLMs are equal (Methods). The fFT order can thus varied *without* overall scaling or additional phases imparted to the field[39], which is critical since we will interfere the beam with its own fFT.

The two-path interferometer in Fig. 2a requires a high degree of stability since several large components (SLMs) are introduced into one path, the overall path lengths may be large (~ 1 m here), and a fractional-transform-order-dependent relative phase must be included (Methods). These difficulties are obviated by introducing a novel configuration that exploits the polarization-selectivity of liquid-crystal-based SLMs[40] to construct the *single-path* polarization interferometer (Fig. 3a). The three SLMs impact the horizontal polarization component H, whereas the vertical component V is unaffected. After rotating the input polarization to 45°, only the H-component is transformed by the SLMs whereas the V-component is unchanged, thus serving as a reference. Projecting the output polarization at 45° allows the H and V components to interfere. However, the V-component undergoes diffraction during propagation and at the output it no longer corresponds to the original field $E(x)$ needed as a reference. We therefore introduce lenses between the SLMs arranged in a $4$-$f$ configuration to image the V-component and reproduce $E(x)$, and modify the strength of the lenses implemented by the SLMs accordingly (Fig. 3b; Methods). Since the symmetry of the configuration is maintained, reflective SLMs allow folding the system such that only two SLMs and one lens are required (Fig. 3c). This stable polarization interferometer is thus in one-to-one correspondence with the two-path interferometer in Fig. 2a.

In implementing the fHT, we require that the SLMs produce simultaneously the equal-order 1D fFTs along $x$ and $y$. Each SLM thus corresponds to equal-power crossed cylindrical lenses, or a spherical lens.

**Results**. We first realize modal analysis via generalized interferometry in the basis of 1D HG modes, where the associated GD is the 1D fFT. We examine beams having the separable form $E(x, y) = E_x(x) E_y(y)$ and focus on the $x$-dependence alone. The input beams are prepared by a single SLM (SLM$_0$) that imprints a phase-only pattern on a Gaussian-mode laser beam, which is then imaged to SLM$_1$ that constitutes the input plane to the generalized interferometer. A second SLM (SLM$_2$) reflects the beam back to SLM$_1$, and the phases imparted by SLM$_1$ and SLM$_2$ are varied to cycle the fFT order $\alpha$.

We report in Fig. 4 measurements carried out on 1D beams approximating the four lowest-order HG modes. For each beam, we provide: (1) the intensity of the 'delayed' beam after the fFT $I(x;\alpha) = |E(x;\alpha)|^2$ while varying the 'delay' $\alpha$; (2) the intensity after interfering the delayed beam with the original, $I_t(x;\alpha) \propto |E(x) + E(x;\alpha)|^2$; (3) the interferogram recorded by the 'bucket detector' $P(\alpha) = \int dx I_t(x;\alpha)$; and (4) the Fourier transform of $P(\alpha)$ that reveals the modal weights $|c_n|^2$. These data enable us to diagnose the system and evaluate its performance, but only the interferogram $P(\alpha)$ is required for modal analysis, which corresponds to the temporal interferogram obtained in traditional two-path interferometers incorporating an optical delay.

Whenever $E(x)$ is a pure $n^{th}$-order HG mode, the interferogram $P(\alpha) \propto 1 + \cos \pi n\alpha/2$ is a sinusoid whose Fourier transform produces a delta function at $n$. We verify this with modes $H_0(x)$ through $H_3(x)$. Because the Gaussian beam $E(x) = H_0(x)$ is an eigenstate of the fFT, we do not observe modulation in $I(x;\alpha)$ or $I_t(x;\alpha)$, and the interferogram $P(\alpha)$ is thus a constant whose Fourier transform has a single contribution at $n = 0$. Next, the 1$^{st}$-order HG mode $E(x) = H_1(x)$ produces an interferogram having a full sinusoidal period $P(\alpha) \propto 1 + \cos \alpha$ whose Fourier transform reveals the strongest contribution at $n = 1$. We approximate $H_1(x)$ by imparting a $\pi$-phase step (via SLM$_0$) to a Gaussian beam, so contributions from other modes appear in the modal analysis, and simulations provide a computed modal content that is in excellent agreement with the measurements. Similarly, $H_2(x)$ and $H_3(x)$ produce shorter period sinusoids and reveal the strongest contributions at $n = 2$ and $n = 3$, respectively. We note a discrepancy at the fFT order $\alpha = 2$, whereupon the rapid variation imposed on the SLM phases results in a sudden drop in diffraction efficiency (Supplementary Information).

We next analyze beams into radial LG modes by implementing the fHT as the GD. The results for $L_0(r^2)$ through $L_2(r^2)$ are presented in Fig. 5. Since these beams are azimuthally invariant, we first integrate the recorded 2D intensity $I(r,\theta)$ in polar coordinates over $\theta$ to obtain a 1D radial distribution $I(r) = \int_0^{2\pi} d\theta\, 2\pi r I(r,\theta)$, where $I(r)$ is the power in a thin annulus of radius $r$ centered on the beam axis. Figure 5a depicts the 'delayed' beam $I(r;\alpha)$ as we vary the fHT-order $\alpha$. Integrating over $r$ after interfering the delayed beam with the reference produces the interferogram $P(\alpha)$. The basis-neutrality is clear when comparing the interferograms associated with $H_0(x)$ in Fig. 4 to $L_0(r^2)$ in Fig. 5; similarly for $H_1(x)$ and $L_1(r^2)$, and for $H_2(x)$ and $L_2(r^2)$.

To highlight the versatility of this approach, we examine beams formed of various superpositions of HG modes in Fig. 6. First, we analyze the beam $E(x) = \{H_0(x) + H_1(x)\}/\sqrt{2}$ which we approximate by blocking half the cross section of a Gaussian beam (Fig. 6a-d). Next, we examine the field $E(x) = \{H_1(x) + iH_2(x)\}/\sqrt{2}$ which we approximate by only varying the phase of a Gaussian beam to maximize the overlap with the desired beam (Fig. 6a-d). Finally, we investigate the superposition $E(x) = \cos\theta\, H_0(x) + i\sin\theta\, H_1(x)$ while varying $\theta$ from 0 to $\pi/2$, thereby switching the beam from $H_0(x)$ to $H_1(x)$ (Fig. 6e-f).

**Discussion and Conclusion**. We have demonstrated that optical interferometry can be generalized to apply for any modal basis by replacing the traditional temporal delay with a generalized delay (GD): an optical transformation that 'delays' the beam in a Hilbert space spanned by the modal basis of interest.

This basis-neutral strategy provides a unifying framework for modal analysis in an arbitrary basis – whether discrete, continuous, or combinations thereof for different degrees of freedom[34]. The fFT performs a rotation of the Wigner distribution associated with the field[43], which has been exploited in tomographically reconstructing the Wigner distribution of non-classical states of light[44]. We have implemented this strategy here in the spatial domain of a scalar field using monochromatic light, but the approach is readily extended to multiple degrees of freedom of the optical field by simply cascading the associated GDs[34]. This methodology is also applicable to quantum states of light, such as one-photon or even entangled two-photon states[45] by replacing the dual delays in a phase-unlocked HOM interferometer[46] with the appropriate GDs. Our approach can thus further increase the accessible dimensionality of the Hilbert space of single photons by at least an order of magnitude[47,48].

The accessible dimension of the beam's Hilbert space is ultimately limited by the spatial resolution of the SLM pixels and the phase-step resolution for each pixel, which limit the sampling resolution of the fractional-transform order. Improvements in SLM technology may allow for real-time modal analysis over large-dimensional Hilbert spaces. One can use instead amplitude-based spatial modulators which are considerably faster, resulting in real-time modal analysis, albeit at the price of reduction in signal throughput[49]. We have found however that the physical extent of the SLM (or the *number* of pixels) is the main factor that limits the fidelity of modal analysis (see Supplementary Information for a detailed study).

Many new question are now open: What is the optimal implementation of a GD when only a closed subspace of the modal basis is of interest? What is the minimum number of SLMs required to implement a GD in an arbitrary modal basis? Moreover, it is usually the case that only a few modes are activated (such as in communications protocols) or contribute significant energy – so-called modal 'sparsity'[50]. In these scenarios, uniformly sampling the GD order is not efficient. We have recently shown theoretically that optical interferometry can be modeled as a linear measurement problem and is hence subject to compressive sensing techniques that exploit the sparsity of the signal in some modal basis[50]. These findings can considerably reduce the number of measurements in the methodology presented here.

We have implemented here the GDs for the Hilbert spaces associated with HG and radial LG modes, the fFT and fHT, respectively. More generally, our approach indicates the potential utility of *yet-to-be-discovered* optical fractional transforms and provides a roadmap for their discovery. Given any modal set of interest, a fractional transform may be constructed out of the outer product of these functions in the diagonal representation given in Eq. 1 – and this fractional transform 'delays' the beam in its associated Hilbert space. For example, one may form a fractional transform from a basis of OAM and Bessel functions for the analysis of beams emerging from optical fibers or circular waveguides.

## Methods

**Properties of a generalized delay**. Consider a functional basis $\{\psi_n(x)\}$ that is orthonormal $\int dx \psi_n^*(x')\psi_m(x) = \delta_{nm}$ and complete $\sum_n \psi_n(x)\psi_n^*(x') = \delta(x-x')$. Using this set as a basis for a 1D finite-energy beam $E(x)$ (in the space of square-integrable functions $L^2$), we have $E(x) = \sum_n c_n \psi_n(x)$, with modal coefficients $c_n = \int dx \psi_n^*(x) E(x)$. For convenience, we normalize the beam energy (the length of a vector in the Hilbert space $L^2$): $\int dx |E(x)|^2 = 1$; consequently, $\sum_n |c_n|^2 = 1$.

Consider a linear, unitary transformation between input and output planes identified by coordinates $x'$ and $x$, respectively. The transformation has a real, continuous order-parameter $\alpha$ that uniquely identifies the transformation $\Lambda(x, x'; \alpha)$. A field $E(x)$ traversing this system is transformed according to Eq. (2), $E(x) \to E(x; \alpha)$. Unitarity implies that $\int dx |E(x)|^2 = \int dx |E(x;\alpha)|^2$ for all $\alpha$ and arbitrary $E(x)$, which implies that

$$K(x', x''; \alpha) = \int dx \Lambda(x, x'; \alpha) \Lambda^*(x, x''; \alpha) = \delta(x' - x''), \forall \alpha. \quad (4)$$

One can thus obtain the form of the GD transformation $\Lambda(x, x'; \alpha)$ in Eq. 1, which further entails that the set of transformations $\Lambda(x, x'; \alpha)$ forms over $\alpha$ a one-parameter group. Defining the group composition operation as the cascade of two transformations, $\Lambda(x, x''; \alpha + \beta) = \int dx' \Lambda(x, x'; \alpha) \Lambda(x', x''; \beta)$, which is closed on this set, we have the requisite properties for a group: (I) the set has an identity $\Lambda(x, x'; 0) = \delta(x - x')$; (II) the group composition operator is associative; and (III) there exists a unique inverse for any transformation $\Lambda(x, x'; \alpha)$, namely $\Lambda(x, x'; -\alpha)$. The group is also obviously commutative. Finally, the property of the inverse and the unitarity of $\Lambda$ together imply that

$$\Lambda(x, x'; -\alpha) = \Lambda^*(x', x; \alpha). \quad (5)$$

**Implementation of the 1D fFT using SLMs**. The 1D fFT is defined by Eq. 1 after substituting the 1D HG functions for $\psi_n(x)$. Explicitly, the 1D fFT is given by the canonical transformation

$$\Lambda_{\text{fFT}}(x, x'; \alpha) = \sqrt{1 - i \cot \alpha} \, \exp\{i\pi(\cot \alpha \, x^2 - 2 \csc \alpha \, xx' + \cot \alpha \, x'^2)\}, \quad (6)$$

where $x$ and $x'$ are normalized and unitless. Several specific angular orders of the fFT are readily recognizable. At $\alpha = 0$, the system is $\Lambda_{\text{fFT}}(x, x'; 0) = \delta(x - x')$, which is an imaging system without inversion or the identity operator; at $\alpha = \frac{\pi}{2}$, $\Lambda_{\text{fFT}}\left(x, x'; \frac{\pi}{2}\right) = \exp\{-i2\pi xx'\}$ is a Fourier transform system; and at $\alpha = \pi$, $\Lambda_{\text{fFT}}(x, x'; \pi) = \delta(x + x')$, which is an imaging system with inversion.

The system in Fig. 3a consists of three cylindrical lenses (implemented by SLMs) of powers $p_1$, $p_2$, and $p_1$ (inverse focal lengths) separated by equal distances $d$, and can perform the 1D fFT of arbitrary order, without scaling or additional spatially varying phase, while using the minimal number of optical components[39]. By introducing a characteristic length scale $\sigma$ (to be set shortly) to normalize $x$ and $x'$, the impulse response function of this system at a wavelength $\lambda$ is

$$h_1(x, x') = \sqrt{\frac{\eta}{2 - p_2 d}} e^{-i\frac{3}{4}\pi} e^{i2kd} \exp\left\{i\pi\eta(x^2 + x'^2)\left(1 - p_1 d - \frac{1}{2 - p_2 d}\right)\right\} \exp\left\{-i\frac{2\pi \eta xx'}{2 - p_2 d}\right\}, \quad (7)$$

where $\eta = \frac{\sigma^2}{\lambda d}$ is a unitless parameter that combines all the length scales in the system. Comparing Eq. 7 to Eq. 6, we identify the lens strengths $p_1$ and $p_2$ that are necessary to implement the fFT of angular order $\alpha$:

$$p_1 d = 1 - \frac{1}{\eta} \cot \frac{\alpha}{2}, \quad p_2 d = 2 - \eta \sin \alpha. \quad (8)$$

In the case of a polarization-selective SLM, the impulse response function for the H-component is Eq. 7 whereas that for the V-component corresponds to free-space propagation for a distance $2d$.

The modified system in Fig. 3b includes two identical lenses with focal lengths $f$ in addition to the three SLMs implementing cylindrical lenses with strengths $s_1$, $s_2$, and $s_1$, and all the separating distances are equal to $f$. The impulse response function for the V-component is $\delta(x+x')$, corresponding to imaging with inversion (a $4f$ imaging system). For the H-component, the impulse response function is a result of all five optical components (three when the system is folded back on itself) is given by:

$$h_2(x,x') = \sqrt{\frac{\eta}{s_2 f}} e^{i\frac{3}{4}\pi} e^{i4kf} \exp\left\{-i\pi\eta(x'^2 + x^2)\left(s_1 f - \frac{1}{s_2 f}\right)\right\} \exp\left\{i\frac{2\pi\eta x x'}{s_2 f}\right\}. \tag{9}$$

where $\eta = \frac{\sigma^2}{\lambda f}$ and we have introduced the transverse length scale $\sigma$ as above. We identify $s_1$ and $s_2$ that implement the 1D fFT of order $\alpha$:

$$s_1 = \frac{1}{\eta f}\cot\frac{\alpha}{2}, \quad s_2 = \frac{\eta}{f}\sin\alpha. \tag{10}$$

This polarization interferometer thus achieves both goals: the H-component undergoes a 1D fFT whereas the V-component reference is imaged via a $4f$ system, both without introducing extra spatial phases or scaling.

**Implementation of the radial fHT using SLMs**. The 2D fFT between input plane $(x',y')$ and output plane $(x,y)$ is separable along the two Cartesian coordinates, such that

$$\Lambda_{xy}(x,y;x',y';\alpha_x,\alpha_x) = \Lambda_x(x,x';\alpha_x)\Lambda_y(y,y';\alpha_y), \tag{11}$$

where $\Lambda_x(x,x';\alpha_x)$ and $\Lambda_y(y,y';\alpha_y)$ are 1D fFTs of order $\alpha_x$ (along $x$) and $\alpha_y$ (along $y$), respectively. These fFTs may be controllably accessed independently by *adding* the phase patterns for the two required crossed generalized cylindrical lenses to be implemented by the SLMs. The fHT corresponds to a symmetric 2D fFT[34,37] $\alpha_x = \alpha_y = \alpha$. In polar coordinates we have $\Lambda_{xy}(x,y;x',y';\alpha,\alpha) \rightarrow \Lambda(r,\theta;r',\theta';\alpha)$. When restricted to azimuthally symmetric functions $E(r,\theta) = E(r)$, $\Lambda$ itself becomes independent of $\theta$ and $\theta'$, $E(r;\alpha) = \int E(r)\Lambda(r,r';\alpha)rdr$, where the purely radial transformation $\Lambda(r,r';\alpha)$ is the fHT, which is thus given by[33]

$$\Lambda(r,r';\alpha) = 2\pi(1-i\cot\alpha)\exp\{i\pi(r'^2+r^2)\cot\alpha\}J_0(2\pi rr'\csc\alpha). \tag{12}$$

Here $J_0(\cdot)$ is the zeroth-order Bessel function of the first kind.

**Experimental setup**. The optical beam is derived from a laser diode at a wavelength of 808 nm that is spatially filtered by coupling into a single-mode fiber at the operating wavelength (Thorlabs, FS-SN-4224) and collimated using a fiber-integrated collimation package. This produces an approximate Gaussian beam whose size is controlled by a variable beam expander (Thorlabs, BE02-05-B) moving along with the collimation package along a rail mount to yield a Gaussian beam with a FWHM of 0.6-mm located at $SLM_0$. The beam is polarized along H and is modulated by $SLM_0$ to produce the desired beam. The field at $SLM_0$ is imaged to $SLM_1$ – through a beam splitter – via a $4f$ imaging system comprised of equal-focal-length lenses ($f = 300$ mm) and the polarization is rotated from H to 45° by a half-wave plate. All the SLMs are reflection-mode, polarization-sensitive Hamamatsu LCOS-SLM (X10468-02) that modulate H but not V. The angle of incidence on $SLM_1$ is less than 10°, the reflected beam passes through a lens $L_1$ ($f = 500$ mm) and is normally incident on $SLM_2$ reflecting back through $L_1$ to $SLM_1$ again. The plane of $SLM_1$ is then imaged to the detector plane through the beam splitter and analyzed at +45° polarization. The image of the modified interference beam is recorded by a CCD camera (The Imaging Source, DFK 72BUC02). The SLMs are computer-controlled to synchronize the display of the phases required to implement the fFT of desired order.

**References**.

[1] N. Kipnis, *History of the Principle of Interference of Light* (Birkhäuser Verlag, 1991).

[2] T. Young, "Experimental demonstration of the general law of the interference of light," *Phil. Trans. Royal Soc. London* **94**, 1-16 (1804).

[3] R. Hanbury Brown and R. Q. Twiss, "Correlation between photons in two coherent beams of light," *Nature* **177**, 27 – 29 (1956).

[4] G. I. Taylor, "Interference fringes with feeble light," *Proc. Cam. Phil. Soc.* **15**, 114-115 (1909).

[5] P. Grangier, G. Roger, and A. Aspect, "Experimental evidence for a photon anticorrelation effect on a beam splitter: A new light on single-photon interferences," *Europhys. Lett.* **1**, 173-179 (1986).

[6] C. K. Hong, Z. Y. Ou, and L. Mandel, "Measurement of subpicosecond time intervals between two photons by interference," *Phys. Rev. Lett.* **59**, 2044-2046 (1987).

[7] D. Malacara, *Optical Shop Testing* (Wiley, New Jersey, 2007) 3$^{rd}$ edition.

[8] D. Huang, *et al*., "Optical coherence tomography," *Science* **254**, 1178-1181 (1991).

[9] B. P. Abbott, *et al*., "GW151226: Observation of gravitational waves from a 22-solar-mass binary black hole coalescence," *Phys. Rev. Lett.* **116**, 241103 (2016).

[10] M. Maldovan and E. L. Thomas, *Periodic Materials and Interference Lithography: for Photonics, Phononics and Mechanics* (Wiley, Weinheim, 2009).

[11] L. Allen, M. W. Beijersbergen, R. J. C. Spreeuw, and J. P. Woerdman, "Orbital angular momentum of light and the transformation of Laguerre-Gaussian laser modes," *Phys. Rev. A* **45**, 8185-8189 (1992).

[12] L. Allen, S. M. Barnett, and M. J. Padgett, *Optical Angular Momentum* (Institute of Physics Publishing, Bristol, 2003).

[13] A. M. Yao and M. J. Padgett, "Orbital angular momentum: origins, behavior and applications," *Adv. Opt. Photon.* **3**, 161-204 (2011).

[14] J. Wang, *et al*., "Terabit free-space data transmission employing orbital angular momentum multiplexing," *Nature Photon.* **6**, 488-496 (2012).

[15] Y. Yan, *et al*., "High-capacity millimetre-wave communications with orbital angular momentum multiplexing," *Nature Commun.* **5**, 4876 (2014).

[16] N. Bozinovic, *et al*., "Terabit-scale orbital angular momentum mode division multiplexing in fibers," *Science* **340**, 1545-1548 (2013).

[17] D. J. Richardson, J. M. Fini, and L. E. Nelson, "Space-division multiplexing in optical fibres," *Nature Photon.* **7**, 354-362 (2013).

[18] A. Mair, A. Vaziri, G. Weihs, and A. Zeilinger, "Entanglement of the orbital angular momentum states of photons," *Nature* **412**, 313-316 (2001).

[19] S. Gröblacher, T. Jennewein, A. Vaziri, G. Weihs, and A. Zeilinger, "Experimental quantum cryptography with qutrits," *New J. Phys.* **8**, 75 (2006).

[20] O. Shapira, A. F. Abouraddy, J. D. Joannopoulos, and Y. Fink, "Complete modal decomposition for optical waveguides," *Phys. Rev. Lett*. **94**, 143902 (2005).

[21] J. W. Nicholson, A. D. Yablon, S. Ramachandran, and S. Ghalmi, "Spatially and spectrally resolved imaging of modal content in large-mode-area fibers," *Opt. Express* **16**, 7233-7243 (2008).

[22] M. Mazilu, A. Mourka, T. Vettenburg, E. M. Wright, and K. Dholakia, "Simultaneous determination of the constituent azimuthal and radial mode indices for light fields possessing orbital angular momentum," *Appl. Phys. Lett.* **100**, 231115 (2012).

[23] A. Dudley, *et al.*, "Efficient sorting of Bessel beams," *Opt. Express* **21**, 165-171 (2013).

[24] M. Stütz, S. Gröblacher, T. Jennewein, and A. Zeilinger, "How to create and detect *N*-dimensional entangled photons with an active-phase hologram," *Appl. Phys. Lett.* **90**, 261114 (2007).

[25] G. C. G. Berkhout, M. P. J. Lavery, J. Courtial, M. W. Beijersbergen, and M. J. Padgett, "Efficient sorting of orbital angular momentum states of light," *Phys. Rev. Lett.* **105**, 153601 (2010).

[26] M. Malik, *et al.*, "Direct measurement of a 27-dimensional orbital-angular-momentum state vector," *Nature Commun.* **5**, 3115 (2014).

[27] A. Forbes, A. Dudley, and M. McLaren, "Creation and detection of optical modes with spatial light modulators," *Adv. Opt. Photon.* **8**, 200-227 (2016).

[28] E. Karimi and E. Santamato, "Radial coherent and intelligent states of paraxial wave equation," *Opt. Lett.* **37**, 2484-2486 (2012).

[29] W. N. Plick, R. Lapkiewicz, S. Ramelow, and A. Zeilinger, "The forgotten quantum number: A short note on the radial modes of Laguerre-Gauss beams," arXiv:1306.6517 (2013).

[30] E. Karimi, *et al.*, "Radial quantum number of Laguerre-Gauss modes," *Phys. Rev. A* **89**, 063813 (2014).

[31] W. N. Plick and M. Krenn, "Physical meaning of the radial index of Laguerre-Gauss beams," *Phys. Rev. A* **92**, 063841(2015).

[32] B. E. A. Saleh and M. C. Teich, *Fundamentals of Photonics* (Wiley, 2007).

[33] A. F. Abouraddy, T. M. Yarnall, and B. E. A. Saleh, "An angular and radial mode analyzer for optical beams," *Opt. Lett.* **36**, 4683-4685 (2011).

[34] A. F. Abouraddy, T. M. Yarnall, and B. E. A. Saleh, "Generalized optical interferometry for modal analysis in arbitrary degrees of freedom," *Opt. Lett.* **37**, 2889-2891 (2012).

[35] V. Namias, "The fractional order Fourier transform and its application to quantum mechanics," *IMA J. Appl. Math.* **25**, 241-265 (1980).

[36] H. M. Ozaktas, Z. Zalevsky, and M. A. Kutay, *The Fractional Fourier Transform* (Wiley, Chisester, 2001).

[37] V. Namias, "Fractionalization of Hankel transforms," *IMA J. Appl. Math.* **26**, 187-197 (1980).

[38] L. Yu, *et al.*, "Deriving the integral representation of a fractional Hankel transform from a fractional Fourier transform," *Opt. Lett.* **23**, 1158-1160 (1998).

[39] J. A. Rodrigo, T. Alieva, and M. L. Calvo, "Programmable two-dimensional optical fractional Fourier processor," *Opt. Express* **17**, 4976-4983 (2009).

[40] K. H. Kagalwala, G. Di Giuseppe, A. F. Abouraddy, and B. E. A. Saleh, "Bell's measure in classical optical coherence," *Nature Photon.* **7**, 72-78 (2013).

[41] A. E. Siegman, *Lasers* (University Science Books, Sausalito, 1986).

[42] A. Sahin, H. M. Ozaktas, and D. Mendlovic, "Optical implementations of two-dimensional fractional Fourier transforms and linear canonical transforms with arbitrary parameters," *Appl. Opt.* **37**, 2130-2141 (1998).

[43] A. W. Lohmann, "Image rotation, Wigner rotation, and the fractional Fourier transform," *J. Opt. Soc. Am. A* **10**, 2181-2186 (1993)

[44] D. T. Smithey, M. Beck, M. G. Raymer, and A. Faridani, "Measurement of the Wigner distribution and the density matrix of a light mode using optical homodyne tomography: Application to squeezed states and the vacuum," *Phys. Rev. Lett.* **70**, 1244-1247 (1993).

[45] A. K. Jahromi, T. M. Yarnall, G. Di Giuseppe, and A. F. Abouraddy, "Hilbert-space analyzers for one-photon and two-photon states," unpublished (2016).

[46] A. F. Abouraddy, T. M. Yarnall, and G. Di Giuseppe, "Phase-unlocked Hong-Ou-Mandel interferometry," *Phys. Rev. A* **87**, 062106 (2013).

[47] W. H. Peeters, E. J. K. Verstegen, and M. P. van Exter, "Orbital angular momentum analysis of high-dimensional entanglement," *Phys. Rev. A* **76**, 042302 (2007).

[48] M. Krenn, *et al*., "Generation and confirmation of a (100×100)-dimensional entangled quantum system," *Proc. Natl. Acad. Sci. USA* **111**, 6122-6123 (2014).

[49] S. Tripathi and K. C. Toussaint, Jr., "Harnessing randomness to control the polarization of light transmitted through highly scattering media," *Opt. Express* **22**, 4412-4422 (2014).

[50] D. Mardani, A. F. Abouraddy, and G. Atia, "Efficient optical mode analysis using compressive interferometry," *Opt. Express* **23**, 28449-28458 (2015).

**Acknowledgments**

We thank T. M. Yarnall, G. Di Giuseppe, B. E. A. Saleh, and D. N. Christodoulides for useful comments. This work was supported by US Office of Naval Research (ONR) contract N00014-14-1-0260.

**Author contributions**

A.F.A developed the concepts. L. M., W. D. L., S.S. performed the optical measurements. D. M., H. E. K., and A. K. J. performed the analysis. H.E.K. prepared the figures. T. M. and A. N. V. carried out additional measurements. A. N. V., G. K. A., and A. F. A. supervised the research. All authors analyzed the data. H. E. K.., D. M., G. K. A., and A.F.A. wrote the paper with input from the co-authors.

**Additional Information**

The authors declare no competing financial interests.

**Figures**

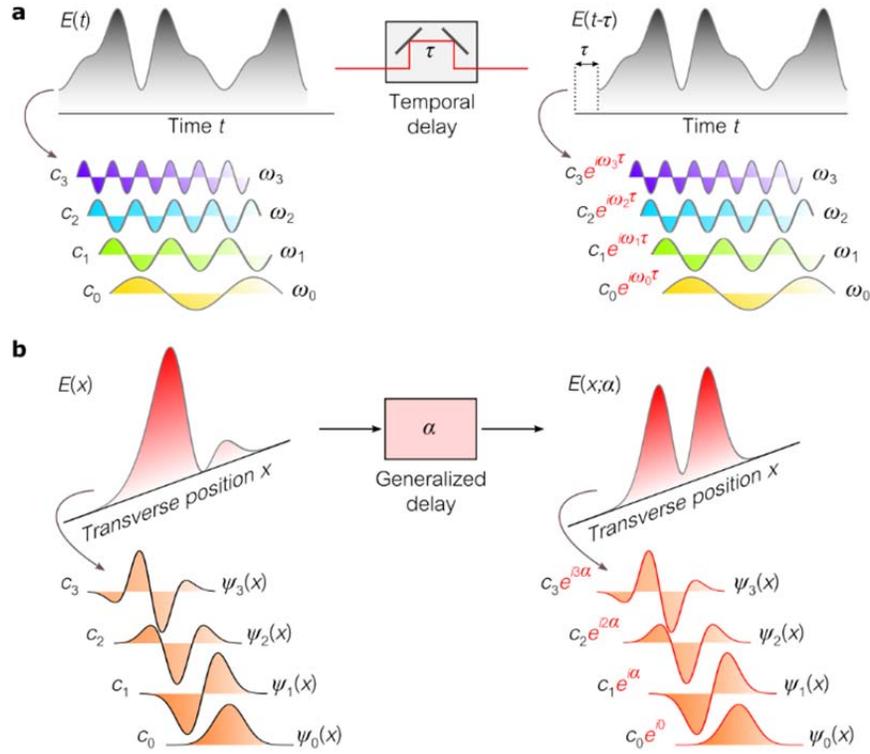

**Figure 1 | Concept of a generalized optical delay. a**, Traditional temporal optical delay. The impact of a temporal delay $\tau$ on a pulse $E(t)$ can be viewed in two ways. In the time domain (first row), the pulse is delayed, $E(t-\tau)$. In the spectral domain (second row), the pulse is a superposition of temporal harmonics $e^{-i\omega t}$ (angular frequencies $\omega$) each with a spectral amplitude $c_n$. The delayed pulse $E(t-\tau)$ is the result of inserting phase factors $e^{i\omega\tau}$ for each harmonic $\omega$. **b**, Generalized delay (GD) $\alpha$ in a Hilbert space spanned by a discrete modal basis $\{\psi_n(x)\}$. The impact of the GD on an optical beam can also be viewed in two domains. In the spatial domain (first row), the GD is not simply a shift but instead it transforms the transverse field profile $E(x) \to E(x;\alpha)$. However, in the modal space (second row) where the field is viewed as a superposition of the modes $\{\psi_n(x)\}$ with weights $c_n$, the impact of the GD is identical to that of the temporal delay on the spectral harmonics in (**a**). The GD adds a phase factor $e^{i\alpha n}$ to the $n^{\text{th}}$ mode amplitude, which 'delays' the beam by $\alpha$ in the Hilbert space spanned by $\{\psi_n(x)\}$.

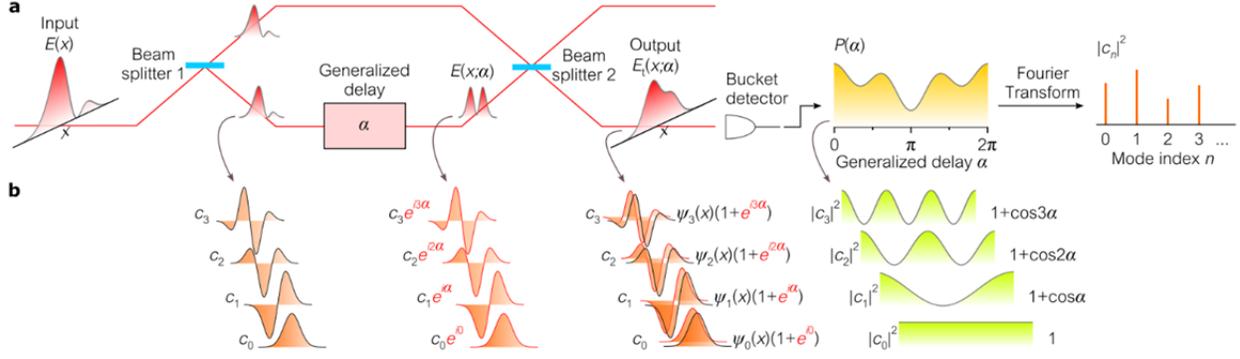

**Figure 2 | Generalized optical interferometry for modal analysis in an arbitrary basis. a**, Operation of a generalized interferometer in real space. Two copies of the beam $E(x)$ are created at beam splitter 1 and subsequently combined at beam splitter 2 after one copy traverses the GD and is 'delayed' in the associated Hilbert space by $\alpha$, $E(x;\alpha)$. The beam emerging from the interferometer – a superposition of the delayed beam and a reference $E_t(x;\alpha) = E(x) + E(x;\alpha)$ – is collected by a bucket detector and an interferogram is recorded with $\alpha$, $P(\alpha) = \int dx |E_t(x;\alpha)|^2$, whose Fourier transform reveals the modal weights $|c_n|^2$. **b**, Operation of the generalized interferometer in the Hilbert space spanned by the modal basis $\{\psi_n(x)\}$ on the beam $E(x) = \sum_n c_n \psi_n(x)$ (Fig. 1b). The underlying modes of the 'delayed' copy acquire phase shifts of the form $e^{in\alpha}$ after passing through the GD to yield a new beam $E(x;\alpha) = \sum_n e^{in\alpha} c_n \psi_n(x)$. The original and 'delayed' beams are combined $E_t(x;\alpha) \propto \sum_n (1 + e^{in\alpha}) c_n \psi_n(x)$ to produce an interferogram $P(\alpha) \propto 1 + \sum_n |c_n|^2 \cos n\alpha$. Because the modes are orthogonal to each other, each interferes only with its phase-shifted counterpart to yield an interferogram of the form $1 + \cos n\alpha$ with weights $|c_n|^2$ – independently of the underlying basis $\{\psi_n(x)\}$ that is traced out at the bucket detector. The sought-after weights are then revealed through harmonic analysis.

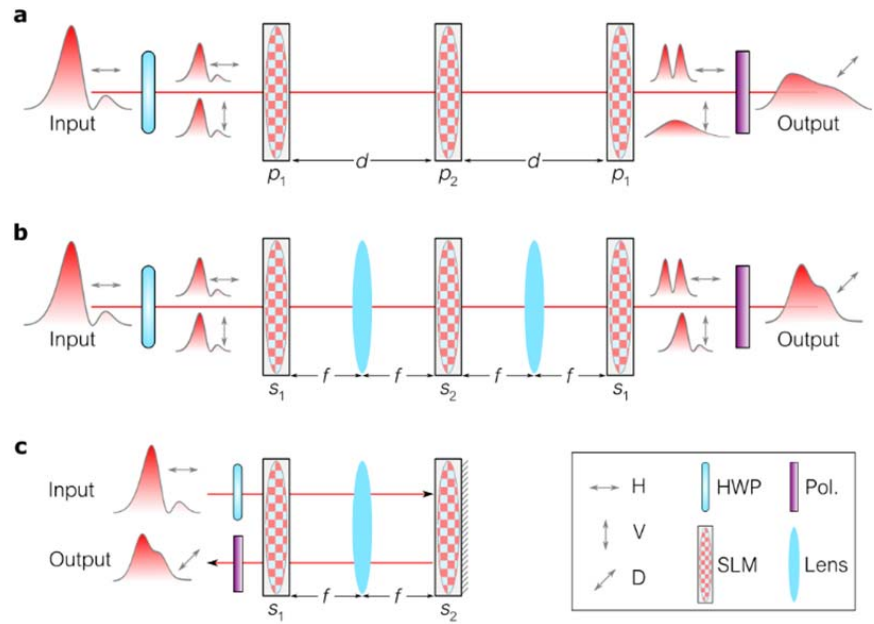

**Figure 3 | Inherently stable implementation of a generalized interferometer. a**, Implementation of a 1D fFT using three generalized (variable-power) lenses $L_1$, $L_2$, and $L_3$ with symmetric strengths $p_1$, $p_2$, and $p_1$, respectively, that are selected to produce a fractional transform of prescribed order (Methods). Because the lenses are implemented by polarization-selective SLMs (affecting only the H-component), the system is in fact equivalent to the two-path interferometer in Fig. 2a, with the H- and V-components corresponding to the delay and reference arms, respectively, while the half-wave plate (HWP) and the polarizer correspond to beam splitters 1 and 2, respectively. This common-path interferometer is inherently stable. However, the V-component undergoes unwanted diffraction over the distance $2d$. **b**, Same as (**a**), except that polarization-insensitive fixed lenses (focal lengths $f$) are inserted in a $4f$ configuration to eliminate the diffraction of the V-component. The strengths $s_1$, $s_2$, and $s_1$ of the generalized lenses are modified to compensate for the added lenses. **c**, Folded implementation of (**b**). The beam is reflected onto itself from $L_2$, such that $L_1$ and $L_3$ are the same generalized lens and only one fixed lens is required.

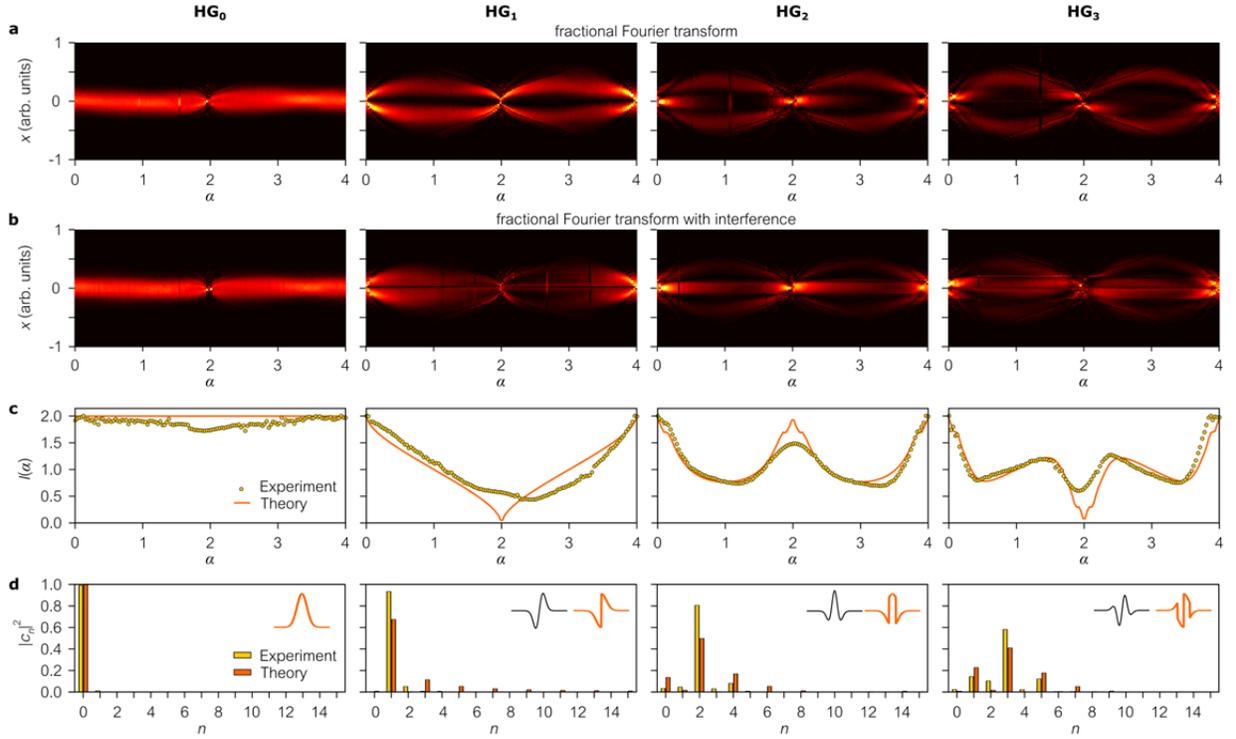

**Figure 4 | Modal analysis in the Hilbert space spanned by 1D Hermite-Gaussian modes using generalized optical interferometry. a**, The measured 'delayed' beam resulting from the input beam $E(x)$ (which is to be analyzed into the contributions from HG modes) traversing the order-$\alpha$ GD (here the fFT), $|E(x;\alpha)|^2$. **b**, The measured interferogram resulting from superposing the delayed beam from (**a**) with a reference, $|E(x) + E(x;\alpha)|^2$. **c**, The integrated interferogram $P(\alpha) = \int dx |E(x) + E(x;\alpha)|^2$. **d**, The modal weights $|c_n|^2$ revealed by taking the Fourier transform of the interferogram in (**c**). The columns are for different input beams corresponding to modes $HG_0$ through $HG_3$. The implemented beams only approximate the pure HG modes (except for $HG_0$ which is exact), as shown in the insets in (**d**). The black mode profile in the inset is an exact HG mode while the orange plot is the approximate beam used in the experiment. The theory plots in (**c**) and (**d**) are those for the implemented approximate beams. See Supplementary Information for theory.

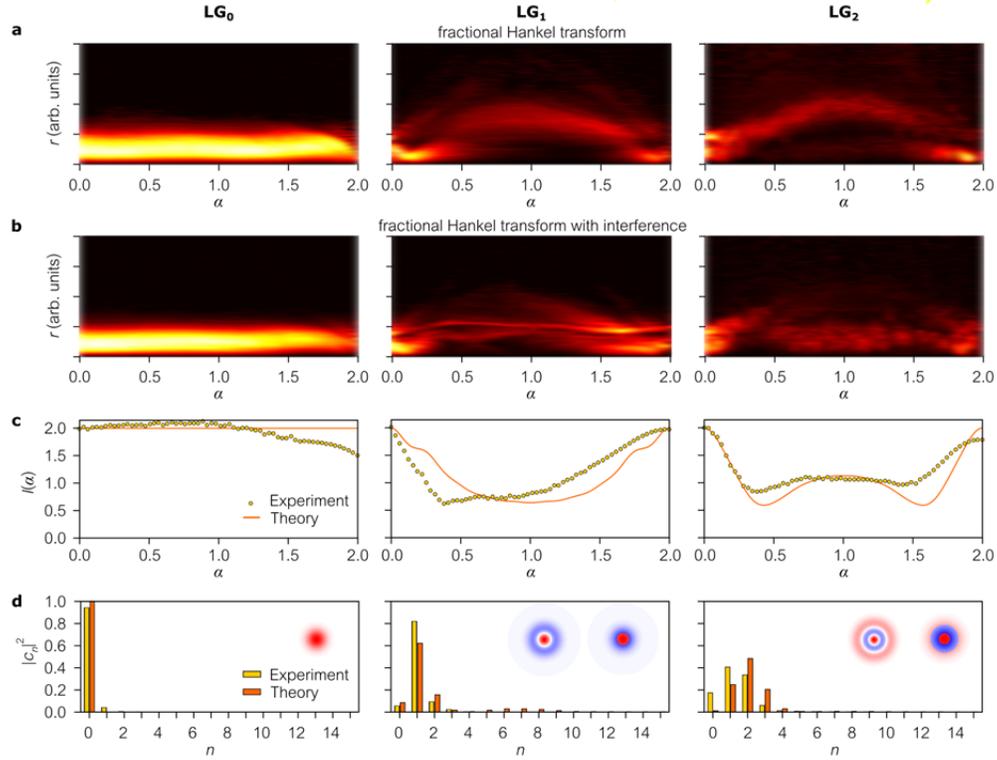

**Figure 5 | Modal analysis in the Hilbert space spanned by radial Laguerre-Gaussian modes using generalized optical interferometry. a-d**, Same as (**a**)-(**d**) in Fig. 4 except that the GD operates in the space of radial LG modes. Note that in (**a**) and (**b**), the delayed beam and the interferogram are plotted with $r$ and not $x$ ($0 \leq r < \infty$). Insets show the radial intensity distribution of the beams. The columns are for different input beams corresponding to modes $LG_0$ through $LG_2$. The implemented beams only approximate the pure radial LG modes (except for $LG_0$ which is exact), as shown in the insets in (**d**). The mode profile on the left in the inset is an exact LG mode while the plot on the right is the approximate beam used in the experiment. The theory plots in (**c**) and (**d**) are those for the implemented approximate beams. See Supplementary Information for theory.

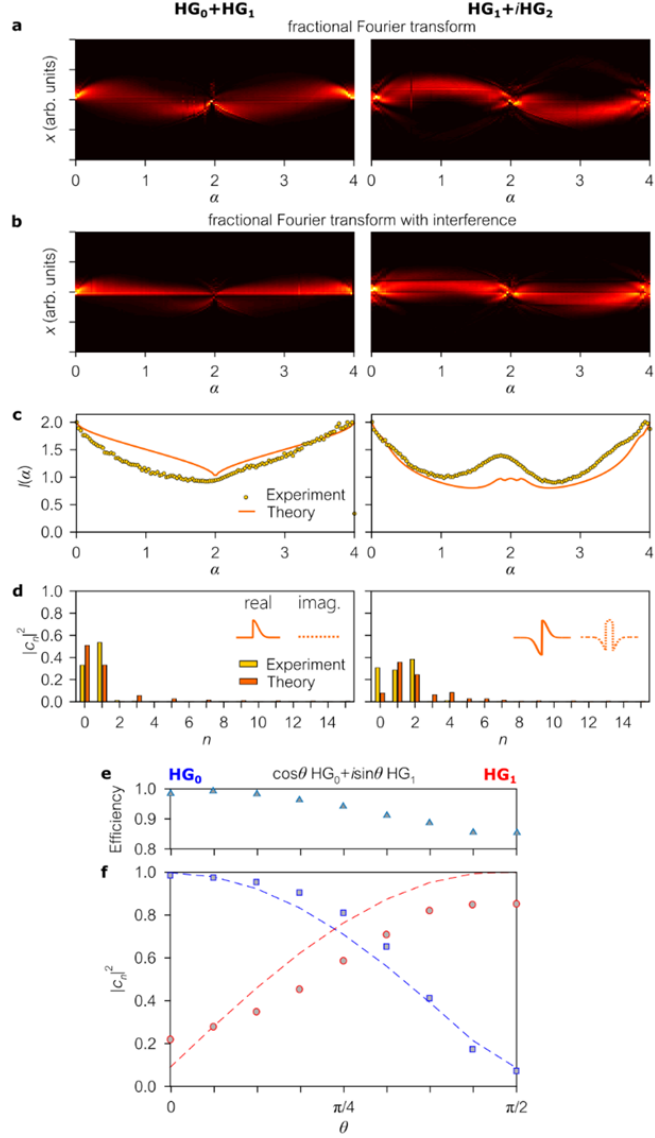

**Figure 6 | Modal analysis of beams comprising superimposed modes. a-d,** Same as (**a**)-(**d**) in Fig. 4 and Fig. 5. The Input beams are the superpositions $H_0(x) + H_1(x)$ (left column) and $H_1(x) + iH_2(x)$ (right column). **e-f,** Modal analysis of the beam $\cos\theta\, H_0(x) + i\sin\theta\, H_1(x)$, while varying $\theta$ from 0 to $\pi/2$. **e**, The efficiency of generating the beam calculated by projecting the vector of experimentally obtained mode amplitudes $|c_n(\theta)|$ onto the vector of theoretically predicted amplitudes $|\hat{c}_n(\theta)|$, $\mathrm{Proj}(\theta) = \sum_n |c_n(\theta)||\hat{c}_n(\theta)|$. **f**, The coefficients $|c_0|^2 = \mathrm{Proj}(\theta, 0)$ (blue squares) and $|c_1|^2 = \mathrm{Proj}(\theta, \pi/2)$ (red circles), corresponding to the contributions of the modes $HG_0$ and $HG_1$. Dashed curves are theoretical predictions, for $HG_0$ we have $\sum_n |\hat{c}_n(\theta)||\hat{c}_n(0)|$ and for $HG_1$ $\sum_n |\hat{c}_n(\theta)||\hat{c}_n(\pi/2)|$.